\documentclass[aps,a4paper,showpacs,%preprint,
groupedaddress]{revtex4}

\usepackage{amsmath,graphicx}

\begin{document}
% You should use BibTeX and apsrev.bst for references
\bibliographystyle{apsrev}

\preprint{TUM-HEP-400/00}

\title{Equivalence between Gaussian averaged neutrino oscillations and
neutrino decoherence}
%\title[]{}

\author{Tommy Ohlsson}
\email[E-mail addresses: ]{tohlsson@physik.tu-muenchen.de, tommy@theophys.kth.se}
\homepage[Homepages: ]{http://www.physik.tu-muenchen.de/~tohlsson/, http://www.theophys.kth.se/~tommy/}
%\thanks{}
%\altaffiliation{}
\affiliation{Institut f{\"u}r Theoretische Physik, Physik Department,
Technische Universit{\"a}t M{\"u}nchen, James-Franck-Stra{\ss}e,
DE-85748 Garching bei M{\"u}nchen, Germany\\ and Division of
Mathematical Physics, Theoretical Physics, Department of Physics,
Royal Institute of Technology, SE-100 44 Stockholm, Sweden}

\date{\today}

\begin{abstract}
In this paper, we show that a Gaussian averaged neutrino oscillation
model is equivalent to a neutrino decoherence model. Without loss of
generality, the analysis is performed with two neutrino flavors.
We also estimate the damping (or decoherence) parameter for
atmospheric neutrinos and compare it to earlier obtained results.
\end{abstract}
% insert suggested PACS numbers in braces on next line

\maketitle

\section{Introduction}

In this paper, we investigate two different models for transitions
of neutrinos and how they are related to each other. The two models
are: neutrino oscillations and neutrino decoherence. Neutrino
oscillation models have been the standard description for neutrino
transitions so far and still are. Neutrino decoherence models, on the
other hand, have recently been discussed by several authors
\cite{sun98,bena00,lisi00,klap00,lisi002,adle00,gago00,gago002} as an
alternative description to neutrino oscillation models. There exist
also other plausible descriptions of neutrino transitions such as
neutrino decay models, which have been suggested by Barger {\it et al.}
\cite{barg99,fogl99,barg992,pakv00}.
However, these models will not be discussed in this paper.

The paper is organized as follows: In Sec.~\ref{sec:form}, we go
through the formalisms of neutrino oscillations and neutrino
decoherence. In Sec.~\ref{sec:equiv}, we give the condition for them
to be equivalent with each other and we also try to estimate the
decoherence term and in Sec.~\ref{sec:est}, we
estimate the damping or decoherence parameter for atmospheric
neutrinos. Finally, in Sec.~\ref{sec:sc}, we present the summary and also
our conclusions.

\section{Formalism}
\label{sec:form}

\subsection{Neutrino oscillations}

The theory of neutrino oscillations is the far most plausible description of
neutrino transitions.
Neutrino oscillations between different neutrino flavors, $\nu_\alpha$
and $\nu_\beta$, occur with the well-known neutrino transition probabilities
\begin{eqnarray}
P_{\alpha\beta} \equiv P_{\alpha\beta}(L,E) = \delta_{\alpha\beta} &-&
4 \; \underset{a < b}{\sum_{a=1}^n \sum_{b=1}^n} \Re (U^\ast_{\alpha
a} U_{\beta a} U_{\alpha b} U^\ast_{\beta b})
\sin^2 \frac{\Delta m_{ab}^2 L}{4E} \nonumber\\
&-& 2 \; \underset{a < b}{\sum_{a=1}^n
\sum_{b=1}^n} \Im (U^\ast_{\alpha a} U_{\beta a}
U_{\alpha b} U^\ast_{\beta b}) \sin \frac{\Delta m_{ab}^2
L}{2E}, \quad \alpha,\beta = e,\mu,\tau,\ldots,
\label{eq:prob_final}
\end{eqnarray}
where $\delta_{\alpha\beta}$ is Kronecker's delta, $L$ is the neutrino
path length, $E$ is the neutrino energy, $n$ is the number of neutrino
flavors, $\Delta m_{ab}^2 \equiv m_a^2
- m_b^2$, $a,b = 1,2,\ldots,n$ is the (vacuum) mass squared difference between
different neutrino mass eigenstates $\nu_a$ and $\nu_b$ (or rather
$\vert \nu_a \rangle$ and $\vert \nu_b \rangle$) with masses $m_a$ and
$m_b$, respectively, and
$$
U = (U_{\alpha a}) = \left( \begin{array}{cccc} U_{e1} & U_{e2} &
U_{e3} & \cdots \\ U_{\mu 1} & U_{\mu 2} & U_{\mu 3} & \cdots \\
U_{\tau 1} & U_{\tau 2} & U_{\tau 3} & \cdots \\ \vdots & \vdots &
\vdots & \ddots \end{array} \right)
$$
is the unitary $n \times n$ Maki--Nakagawa--Sakata (MNS)
mixing matrix \cite{maki62}, which relates the flavor states
$\vert \nu_\alpha \rangle$ ($\alpha = e,\mu,\tau,\ldots$)
and the mass eigenstates $\vert \nu_a \rangle$ ($a = 1,2,\ldots,n$).

However, since in practice a neutrino wave is neither detected nor produced
with sharp energy or with well-defined propagation length, we have to average
over the $L/E$ dependence and other uncertainties in the detection and
emission of the neutrino wave.

We will here use the Gaussian average, which is defined by
\begin{equation}
\langle P \rangle \equiv \int_{-\infty}^{\infty} P(x) f(x) \, dx,
\end{equation}
where
$$
f(x) \equiv \frac{1}{\sigma \sqrt{2 \pi}} e^{-\frac{(x-\ell)^2}{2 \sigma^2}}.
$$
Here $\ell \equiv \langle x \rangle$ and $\sigma \equiv \sqrt{\langle
(x - \langle x \rangle)^2 \rangle}$ are the expectation value and standard
deviation, respectively.

By taking the Gaussian average of Eq.~(\ref{eq:prob_final}) and using
$x \equiv \frac{L}{4E}$, we obtain the averaged transition
probabilities from $\nu_\alpha$ to $\nu_\beta$ as
\begin{eqnarray}
\langle P_{\alpha\beta} \rangle = \delta_{\alpha\beta} &-& 2 \;
\underset{a < b}{\sum_{a=1}^n \sum_{b=1}^n} \Re(U_{\alpha a}^\ast U_{\beta a}
U_{\alpha b} U_{\beta b}^\ast) \left( 1- \cos \left(2 \ell \Delta
m_{ab}^2\right) e^{-2 \sigma^2 (\Delta m_{ab}^2)^2} \right) \nonumber\\
&-& 2 \; \underset{a < b}{\sum_{a=1}^n \sum_{b=1}^n} \Im(U_{\alpha
a}^\ast U_{\beta a} U_{\alpha b} U_{\beta b}^\ast) \sin \left(2 \ell \Delta
m_{ab}^2\right) e^{-2 \sigma^2 (\Delta m_{ab}^2)^2}, \quad 
\alpha,\beta = e, \mu, \tau, \ldots.
\label{eq:<Pab>}
\end{eqnarray}
The physical interpretations of the parameters $\ell$ and $\sigma$
are the following:
\begin{itemize}
\item {\bf The parameter $\mbox{\boldmath$\ell$}$:} The parameter
$\ell$ deals with the sensitivity of an experiment and is given by
$\ell \equiv \langle L/E \rangle/4$. Note that we will here use
$\langle L/E \rangle = \langle L \rangle / \langle E \rangle$, {\it
i.e.},
\begin{equation}
\ell = \frac{\langle L \rangle}{4 \langle E \rangle}.
\end{equation}
This simplification holds if $L$ and $E$ are independent.
\item {\bf The parameter $\mbox{\boldmath$\sigma$}$:} The parameter
$\sigma$ is a so-called damping factor, which is responsible for the
damping of the neutrino oscillation probabilities. 

A pessimistic upper bound for the damping parameter $\sigma$ is given by the
uncertainty in $x$, {\it i.e.},
\begin{equation}
\sigma \simeq \Delta x = \Delta \frac{L}{4E} \leq \left\vert
\frac{\partial x}{\partial L} \right\vert_{L = \langle L \rangle, E =
\langle E \rangle} \Delta L + \left\vert
\frac{\partial x}{\partial E} \right\vert_{L = \langle L \rangle, E =
\langle E \rangle} \Delta E = \frac{\langle L \rangle}{4 \langle E
\rangle} \left( \frac{\Delta L}{\langle L \rangle} + \frac{\Delta
E}{\langle E \rangle} \right),
\label{eq:sigma}
\end{equation}
where $\Delta L$ and $\Delta E$ are the uncertainties in the neutrino
path length and the neutrino energy, respectively. A more optimistic
upper bound would be
\begin{equation}
\sigma \lesssim \frac{\langle L \rangle}{4 \langle E \rangle}
\sqrt{\left(\frac{\Delta L}{\langle L \rangle}\right)^2 +
\left(\frac{\Delta E}{\langle E \rangle}\right)^2}.
\end{equation}

For large values of $\sigma$, the dependence on the mass squared
differences will be completely washed out, since $1 - \cos \left(2
\ell \Delta m_{ab}^2\right) e^{-2 \sigma^2 (\Delta m_{ab}^2)^2} \to 1$
and $\sin \left(2 \ell \Delta m_{ab}^2\right) e^{-2 \sigma^2 (\Delta
m_{ab}^2)^2} \to 0$ when $\sigma \to \infty$, and the Gaussian
averaged transition probabilities $\langle P_{\alpha \beta} \rangle$
will just be dependent on the MNS mixing matrix elements, the $U_{\alpha
a}$'s, {\it i.e.},
\begin{equation}
\lim_{\sigma \to \infty} \langle P_{\alpha\beta} \rangle =
\delta_{\alpha\beta} - 2 \;
\underset{a < b}{\sum_{a=1}^n \sum_{b=1}^n} \Re(U_{\alpha a}^\ast U_{\beta a}
U_{\alpha b} U_{\beta b}^\ast).
\label{eq:00<Pab>}
\end{equation}
Note that the imaginary part sum (the second sum in Eq.~(\ref{eq:<Pab>}))
does not appear at all (in any form) in
Eq.~(\ref{eq:00<Pab>}). Equation~(\ref{eq:00<Pab>}) corresponds to the  
classical limit.

In the other limit, $\sigma \to 0$,
we will just regain Eq.~(\ref{eq:prob_final}) from
Eq.~(\ref{eq:<Pab>}) with $\ell = x$, {\it i.e.},
\begin{equation}
\lim_{\sigma \to 0} \langle P_{\alpha \beta} \rangle = P_{\alpha
\beta}.
\label{eq:0<Pab>}
\end{equation}
\end{itemize}

In the case of two neutrino flavors ($n = 2$), we can call these
neutrino flavors $\nu_e$ and $\nu_\mu$, the unitary (orthogonal) $2
\times 2$ MNS mixing matrix $U$ is usually parameterized as
\begin{equation}
U = (U_{\alpha a}) = \left( \begin{array}{cc} U_{e1} & U_{e2} \\
U_{\mu 1} & U_{\mu 2} \end{array} \right) = \left( \begin{array}{cc}
\cos \theta & \sin \theta \\ -\sin \theta & \cos \theta \end{array} \right),
\label{eq:U2}
\end{equation}
where $\theta$ is the (vacuum) mixing angle. Note that $U^\ast = U$ in
this case. Furthermore, $\Delta m^2 \equiv \Delta m^2_{21} = - \Delta
m^2_{12}$. Observe that for two neutrino flavors there cannot be any ${\cal
CP}$ (or ${\cal T}$) violation in the MNS mixing matrix $U$, since $U$
is a real matrix in this case, which means that
\begin{equation}
P_{ee} = 1 - P_{e\mu} = 1 - P_{\mu e} = P_{\mu\mu}.
\label{eq:peepmm}
\end{equation}
By taking the average of Eq.~(\ref{eq:peepmm}), we of course also have
\begin{equation}
\langle P_{ee} \rangle = 1 - \langle P_{e \mu} \rangle = 1 - \langle
P_{\mu e} \rangle = \langle P_{\mu\mu} \rangle.
\end{equation}
Thus, inserting $\alpha = e$, $\beta = \mu$, and $n = 2$ into
Eq.~(\ref{eq:<Pab>}), we obtain the Gaussian averaged two flavor
neutrino oscillation formula as
\begin{equation}
\langle P_{e \mu} \rangle (\ell) = \frac{1}{2} \sin^2 2\theta \left( 1 - e^{-
2 \sigma^2 \left( \Delta m^2 \right)^2} \cos \left( 2 \ell \Delta m^2
\right) \right),
\end{equation}
where $\ell \simeq \langle L \rangle /(4 \langle E \rangle)$, {\it
i.e.},
\begin{equation}
\langle P_{e \mu} \rangle (\langle L \rangle, \langle E \rangle) =
\frac{1}{2} \sin^2 2\theta \left( 1 - e^{- 2 \sigma^2 \left( \Delta
m^2 \right)^2} \cos \frac{\Delta m^2 \langle L \rangle}{2 \langle E
\rangle} \right).
\label{eq:neuosc}
\end{equation}
In what follows, we will use $\langle P_{\alpha \beta} \rangle =
P_{\alpha \beta}$, $\langle L \rangle = L$, and $\langle E \rangle =
E$ for simplicity.

\subsection{Neutrino decoherence}

Neutrino decoherence arises when we consider a neutrino system to be
coupled to an environment (or a reservoir or a bath). In general, we
are not allowed to use the Schr{\"o}dinger equation to describe
system-environment type interactions, since even if we start initially
with a pure quantum mechanical state, the coupling to the environment
will produce mixed quantum mechanical states. Thus, we are forced to
use the Liouville equation.

Assume that $\rho = \rho(t)$ is the neutrino density
matrix \footnote{The complete time evolution of the neutrino density
matrix $\rho$ is given by a quantum mechanical semigroup, {\it i.e.},
by a completely positive, trace-preserving family of linear maps:
$\gamma_t: \rho(0) \mapsto \rho(t)$ \cite{bena00}.}, which is Hermitian
($\rho^\dagger = \rho$) with positive eigenvalues and has constant
trace equal to one (${\rm tr \,} \rho = 1$). The Liouville equation
for the neutrino density matrix is then
\begin{equation}
\dot{\rho} = - i \left[H_m,\rho\right],
\label{eq:liou}
\end{equation}
where $H_m$ is the free Hamiltonian for the neutrinos in the mass eigenstate
basis. We will assume that the neutrino system is a two-level system,
{\it i.e.}, we have two neutrino flavors ($n = 2$), $\nu_e$ and
$\nu_\mu$, as in the neutrino oscillation formalism. Thus, in this
case, the neutrino density matrix $\rho$ is a $2 \times 2$ matrix and
the free Hamiltonian $H_m$ is also a $2 \times 2$ matrix and is given by
\begin{equation}
H_m = \frac{1}{2 E}{\rm diag \,}(m_1^2,m_2^2) \quad \mbox{or} \quad H_m
\mapsto H_m' = H_m - \frac{1}{2} ({\rm tr \,} H_m) I_2 = \frac{1}{4 E}
{\rm diag \,}(-\Delta m^2,\Delta m^2),
\end{equation}
where $I_2$ is the $2 \times 2$ identity matrix and again $\Delta m^2
\equiv m_2^2 - m_1^2$. The traceless Hamiltonian $H_m'$ can be used
instead of the Hamiltonian $H_m$, since they just differ by a trace
(of any of them). This trace will only give rise to a global phase
that will not affect the transition probabilities anyway and is
therefore irrelevant. 

Furthermore, note that we will not consider matter effects in this
paper. In the case of neutrino transitions in matter, the Hamiltonian
has a much more complicated structure.

Solving Eq.~(\ref{eq:liou}) under the above assumptions leads to the
ordinary neutrino oscillation transition probabilities in
Eq.~(\ref{eq:prob_final}) with $n = 2$ as it should. However, the
neutrinos could be influenced by so-called decoherence effects. Such
effects are introduced by an extra term ${\cal D}[\rho]$ in the
Liouville equation for the neutrinos, {\it i.e.}, the Markovian
Liouville--Lindblad quantum mechanical master equation
\cite{lind76,alic87,bena00}
\begin{equation}
\dot{\rho} = - i \left[H_m,\rho\right] - {\cal D}[\rho],
\label{eq:liou_lind}
\end{equation}
which allows transitions from pure quantum mechanical states to mixed
quantum mechanical states. We will here use the Lindblad form for the
decoherence term \cite{lind76}, {\it viz.},
\begin{equation}
{\cal D}[\rho] = \sum_{a = 1}^n \left( \{ \rho, D_a^\dagger D_a \} - 2
D_a \rho D_a^\dagger \right),
\label{eq:gli}
\end{equation}
where the $D_a$'s are Lindblad operators \footnote{Statements by
Adler \cite{adle00}:\\ 1. For a two-level system, the only choices of
$D_a$ that commute with $H_m$ are either $D_a = \delta_a I_2$
(trivial, since it makes no contribution to the decoherence term
${\cal D}[\rho]$ [Eq.~(\ref{eq:gli2})] and it can therefore be ignored)
or $D_a = d_a H_m$.\\ 2. The sum in the decoherence term can be replaced
with a single Lindblad operator $D = d H_m$.} arising from tracing or averaging
away environment dynamics and must be such that $\sum_{a = 1}^n
D_a^\dagger D_a$ is a well-defined $2 \times 2$ matrix. Note that the
extra term ${\cal D}[\rho]$ is responsible for the fact that the
quantum mechanical states can develop dissipation and irreversibility,
and possible loss of quantum coherence {\it i.e.}, (quantum)
decoherence \cite{elli84,bena00}. The $D_a$'s are normally Hermitian
($D_a^\dagger = D_a$) if we require monotone time increase of the von
Neumann entropy, $S = - {\rm tr \,}(\rho \log \rho)$, and they usually
also commute with the Hamiltonian ($\left[H_m,D_a\right] = 0$) if we
require conservation of the statistical average of the energy, {\it i.e.},
$\frac{d}{dt} {\rm tr \,}(H_m \rho) = 0$.

Assuming that the $D_a$'s are Hermitian, Eq.~(\ref{eq:gli}) becomes
\begin{equation}
{\cal D}[\rho] = \sum_{a = 1}^n \left[D_a, \left[D_a, \rho\right]
\right] = \sum_{a = 1}^n \left( \rho D_a^2 + D_a^2 \rho - 2 D_a \rho
D_a \right).
\label{eq:gli2}
\end{equation}
The operators $\rho$, $H_m$, $D_a$, which are all Hermitian, can therefore
be expanded in the Pauli matrix basis as
\begin{eqnarray}
\rho &=& \frac{1}{2} \left( I_2 + {\bf p} \cdot \mbox{\boldmath$\sigma$}
\right), \\
H_m &=& \frac{1}{2} {\bf k} \cdot \mbox{\boldmath$\sigma$}, \\
D_a &=& \frac{1}{2} {\bf d}_a \cdot \mbox{\boldmath$\sigma$}, \quad a
= 1,2,\ldots,n,
\end{eqnarray}
where $I_2$ is again the $2 \times 2$ identity matrix,
$\mbox{\boldmath$\sigma$} \equiv (\sigma_1,\sigma_2,\sigma_3)$ is the Pauli
matrix vector, and ${\bf k} \equiv (0,0,-k)$. Here $k \equiv \frac{\Delta
m^2}{2E}$. In the Pauli matrix basis, we have after some tedious calculations
\begin{eqnarray}
\dot{\rho} &=& \frac{1}{2} \dot{{\bf p}} \cdot
\mbox{\boldmath$\sigma$}, \\
\left[H_m,\rho\right] &=& \frac{i}{2} \left( {\bf k} \times {\bf p}
\right) \cdot \mbox{\boldmath$\sigma$}, \\
{\cal D}[\rho] &=& \frac{1}{2} \sum_{a = 1}^n \left( \vert {\bf d}_a
\vert^2 {\bf p} \cdot \mbox{\boldmath$\sigma$} - ({\bf d}_a \cdot {\bf
p}) ({\bf d}_a \cdot \mbox{\boldmath$\sigma$}) \right),
\end{eqnarray}
which means that Eq.~(\ref{eq:liou_lind}) can be written in the form
\begin{equation}
\dot{{\bf p}} \cdot \mbox{\boldmath$\sigma$} = \left( {\bf k} \times
{\bf p} \right) \cdot \mbox{\boldmath$\sigma$} - \sum_{a = 1}^n \left(
\vert {\bf d}_a \vert^2 {\bf p} \cdot \mbox{\boldmath$\sigma$} - ({\bf
d}_a \cdot {\bf p})({\bf d}_a \cdot \mbox{\boldmath$\sigma$}) \right).
\end{equation}
Using well-known formulas from vector algebra, the sum in the above
equation can be expressed as
$$
\sum_{a = 1}^n \left( {\bf d}_a \times \left( {\bf d}_a \times {\bf p}
\right) \right) \cdot \mbox{\boldmath$\sigma$}.
$$
Thus, we have
\begin{equation}
\dot{{\bf p}} \cdot \mbox{\boldmath$\sigma$} = \left( {\bf k} \times
{\bf p} \right) \cdot \mbox{\boldmath$\sigma$} - \sum_{a = 1}^n \left(
{\bf d}_a \times \left( {\bf d}_a \times {\bf p} \right) \right) \cdot
\mbox{\boldmath$\sigma$}
\end{equation}
or
\begin{equation}
\dot{{\bf p}} = {\bf k} \times {\bf p} - \sum_{a = 1}^n {\bf d}_a
\times \left( {\bf d}_a \times {\bf p} \right) = {\bf k} \times {\bf
p} - \left( \sum_{a = 1}^n \vert {\bf d}_a \vert^2 \right) {\bf p} +
\sum_{a = 1}^n ({\bf d}_a \cdot {\bf p}) {\bf d}_a,
\label{eq:liou_lind_vec}
\end{equation}
which is the Bloch--Lindblad equation. The first term after the second equality
sign in the Bloch--Lindblad equation, ${\bf k} \times {\bf p}$, is giving rise
to neutrino oscillations, whereas the second and third terms are the
decoherence terms.

If $\left[H_m,D_a\right] = 0$, then it follows that ${\bf k} \times {\bf d}_a
= {\bf 0}$, {\it i.e.}, ${\bf k}$ and ${\bf d}_a$ are parallel
vectors. Thus, we can put ${\bf d}_a = d_a \hat{{\bf k}}$, where
$\hat{{\bf k}} \equiv {\bf k}/\vert {\bf k} \vert$. Inserting this
into Eq.~(\ref{eq:liou_lind_vec}), we obtain
\begin{equation}
(\dot{p}_1,\dot{p}_2,\dot{p}_3) = (kp_2,-kp_1,0) - d^2
(p_1,p_2,p_3) + (0,0,d^2 p_3),
\end{equation}
where $d^2 \equiv \sum_{a = 1}^n d_a^2$, {\it i.e.},
\begin{eqnarray}
\dot{p}_1 &=& k p_2 - d^2 p_1, \\
\dot{p}_2 &=& - k p_1 - d^2 p_2, \\
\dot{p}_3 &=& 0,
\end{eqnarray}
which can be written in matrix form as ($\dot{{\bf p}} = M {\bf p}$)
\begin{equation}
\left( \begin{array}{c} \dot{p}_1 \\ \dot{p}_2 \\ \dot{p}_3
\end{array} \right) = \left( \begin{array}{ccc} - d^2 & k & 0
\\ - k & - d^2 & 0 \\ 0 & 0 & 0 \end{array} \right) \left(
\begin{array}{c} p_1 \\ p_2 \\ p_3 \end{array} \right).
\end{equation}
This system of first order differential equations has the solution
\begin{equation}
{\bf p}(t) \equiv \left( \begin{array}{c} p_1(t) \\
p_2(t) \\ p_3(t) \end{array} \right) = \left( \begin{array}{ccc}
e^{-d^2 t} \cos kt & e^{-d^2 t} \sin kt & 0 \\ -e^{-d^2 t} \sin kt &
e^{-d^2 t} \cos kt & 0 \\ 0 & 0 & 1
\end{array} \right) \left( \begin{array}{c} p_1(0) \\ p_2(0) \\ p_3(0)
\end{array} \right) \equiv e^{Mt} {\bf p}(0),
\end{equation}
{\it i.e.},
\begin{eqnarray}
p_1(t) &=& p_1(0) e^{-d^2 t} \cos kt + p_2(0) e^{-d^2 t} \sin kt,
\label{eq:p1t} \\
p_2(t) &=& - p_1(0) e^{-d^2 t} \sin kt + p_2(0) e^{-d^2 t} \cos kt, \\
p_3(t) &=& p_3(0). \label{eq:p3t}
\end{eqnarray}

Another, but equivalent, way of obtaining Eq.~(\ref{eq:liou_lind_vec})
has been found by Stodolsky. Stodolsky's formula is given by
\cite{stod87,raff96}
\begin{equation}
\dot{{\bf P}} = {\bf V} \times {\bf P} - D {\bf P}_{\rm T},
\label{eq:stod}
\end{equation}
where ${\bf P}$ is the ``polarization vector'', ${\bf V}$ is the
``magnetic field'', $D$ is the damping parameter determined by the
scattering amplitudes on the background ({\it i.e.}, the environment),
and ${\bf P}_{\rm T}$ is the ``transverse'' part of ${\bf P}$. In a
two-level system, the length of ${\bf P}$ measures the degree of
coherence: $\vert {\bf P} \vert = 1$ corresponds to a pure state, $0 <
\vert {\bf P} \vert < 1$ to some degree of incoherence, and $\vert
{\bf P} \vert = 0$ to the completely mixed or incoherent state
\cite{stod87}, {\it i.e.}, the loss of coherence is given by the
shrinking of $\vert {\bf P} \vert$. The time evolution described by 
Eq.~(\ref{eq:stod}) is a precession around ${\bf V}$, combined with
the shrinking of $\vert {\bf P} \vert$ to zero. The final state
corresponds to $\rho = \frac{1}{2}$, where both neutrino flavor states,
$\nu_e$ and $\nu_\mu$, are equally populated, and with vanishing
coherence between them. In our notation, ${\bf P} \equiv {\bf p}$,
${\bf V} \equiv {\bf k}$, and $D {\bf p}_{\rm T} \equiv \left( \sum_{a
= 1}^n \vert {\bf d}_a \vert^2 \right) {\bf p} - \sum_{a = 1}^n ({\bf
d}_a \cdot {\bf p}) {\bf d}_a = d^2 ( {\bf p} - ({\bf p} \cdot
\hat{{\bf k}}) \hat{{\bf k}})$. Since here $\hat{{\bf k}} = - {\bf
e}_z$, we have that $D {\bf P}_{\rm T} = d^2 ( {\bf p} - p_3 {\bf e}_z
) = d^2 (p_1,p_2,0)$.

Thus, we can now construct the time-dependent neutrino density matrix as
\begin{equation}
\rho(t) = \frac{1}{2} \left( I_2 + {\bf p}(t) \cdot
\mbox{\boldmath$\sigma$} \right) = \frac{1}{2} \left( \begin{array}{cc} 1 +
p_3(t) & p_1(t) - i p_2(t) \\ p_1(t) + i p_2(t) & 1 - p_3(t)
\end{array} \right).
\label{eq:rho}
\end{equation}

As usual, the neutrino flavor states $\vert \nu_\alpha \rangle$,
where $\alpha = e,\mu$, are superpositions of the neutrino mass eigenstates
$\vert \nu_a \rangle$, where $a = 1,2$, {\it i.e.}, 
\begin{equation}
\vert \nu_\alpha \rangle = \sum_{a = 1}^2 U_{\alpha a}^\ast \vert \nu_a
\rangle = \sum_{a = 1}^2 U_{\alpha a} \vert \nu_a \rangle, \quad
\alpha = e,\mu.
\end{equation}
If we represent the mass eigenstates by the vectors $\vert \nu_1
\rangle = (1,0)$ and $\vert \nu_2 \rangle = (0,1)$, then the flavor
states become $\vert \nu_e \rangle = (\cos \theta, \sin \theta)$ and
$\vert \nu_\mu \rangle = (-\sin \theta, \cos \theta)$. Furthermore, if
we assume that the initial state of a neutrino is $\vert \nu_e
\rangle$, {\it i.e.}, the system is prepared in a pure $\nu_e$ state,
then the initial condition for the neutrino density matrix is
\begin{equation}
\rho(0) \equiv \rho_{ee} \equiv \vert \nu_e \rangle \langle \nu_e
\vert = \left( \begin{array}{cc} \cos^2 \theta & \sin \theta \cos
\theta \\ \sin \theta \cos \theta & \cos^2 \theta \end{array} \right)
= \frac{1}{2} \left( I_2 + {\bf p}(0) \cdot \mbox{\boldmath$\sigma$} \right),
\end{equation}
and it follows, using Eq.~(\ref{eq:rho}), that $p_1(0) = \sin
2\theta$, $p_2(0) = 0$, and $p_3(0) = \cos 2\theta$.
Similarly, if the initial state is $\vert \nu_\mu \rangle$, then we
have
\begin{equation}
\rho_{\mu\mu} \equiv \vert \nu_\mu \rangle \langle \nu_\mu \vert =
\left( \begin{array}{cc} \sin^2 \theta & -\sin \theta \cos \theta \\
-\sin \theta \cos \theta & \cos^2 \theta \end{array} \right) = 1 - \rho_{ee}.
\end{equation}

Thus, the neutrino transition probabilities with decoherence effects read
\begin{eqnarray}
P_{e e}(t) &\equiv& {\rm tr \,}(\vert \nu_e \rangle \langle \nu_e \vert
\rho(t)) = \frac{1}{2} \left[ 1 + \sin 2\theta p_1(t) + \cos 2\theta
p_3(t) \right], \label{eq:Pee} \\
P_{e \mu}(t) &\equiv& {\rm tr \,}(\vert \nu_\mu \rangle \langle
\nu_\mu \vert \rho(t)) = \frac{1}{2} \left[1 - \sin 2\theta p_1(t) -
\cos 2\theta p_3(t) \right].
\label{eq:Pem}
\end{eqnarray}
Moreover, as in the neutrino oscillation formalism, we have the
relation between the transition probabilities
\begin{equation}
P_{e \mu}(t) = 1 - P_{e e}(t) = 1 - P_{\mu \mu}(t) = P_{\mu e}(t),
\end{equation}
since
\begin{equation}
1 = {\rm tr \,} \rho(t) = {\rm tr \,}(1 \rho(t)) = {\rm tr \,}\left(
\sum_{\alpha = e,\mu} \vert \nu_\alpha \rangle \langle \nu_\alpha
\vert \rho(t)\right) = {\rm tr \,}(\vert \nu_e \rangle \langle \nu_e \vert
\rho(t)) + {\rm tr \,}(\vert \nu_\mu \rangle \langle \nu_\mu \vert
\rho(t)) = P_{ee}(t) + P_{e \mu}(t).
\end{equation}
Inserting Eqs.~(\ref{eq:p1t}) - (\ref{eq:p3t}) with the above initial
conditions into Eq.~(\ref{eq:Pem}) gives
\begin{equation}
P_{e\mu}(t) = \frac{1}{2} \sin^2 2\theta \left( 1 - e^{-d^2 t}
\cos kt \right).
\end{equation}
Furthermore, inserting $t \simeq L$ and $k = \Delta m^2/(2E)$, we thus
obtain the neutrino decoherence formula for two neutrino flavors as
\begin{equation}
P_{e \mu}(L,E) = \frac{1}{2} \sin^2 2\theta \left( 1 - e^{-d^2 L} \cos
\frac{\Delta m^2 L}{2E} \right).
\label{eq:neudeco}
\end{equation}

When $\Delta m^2 = 0$ (no neutrino oscillations), we find the pure
neutrino decoherence formula
\begin{equation}
P_{e \mu}(L) = \frac{1}{2} \sin^2 2\theta \left( 1 - e^{-d^2 L} \right).
\end{equation}
Note that this formula is explicitly independent of the neutrino energy $E$.
Other interesting limits are obtained when $d = 0$ (no neutrino
decoherence)
\begin{equation}
P_{e \mu}(L,E) = \frac{1}{2} \sin^2 2\theta \left( 1 - \cos
\frac{\Delta m^2 L}{2 E} \right),
\end{equation}
{\it i.e.}, the well-known Pontecorvo neutrino oscillation formula
({\it cf.} Eq.~(\ref{eq:0<Pab>})) and
when $d \to \infty$ (or $\Delta m^2 \gg 2E/L$)
\begin{equation}
P_{e \mu} = \frac{1}{2} \sin^2 2\theta,
\end{equation}
{\it i.e.}, the classical transition probability formula ({\it cf.}
Eq.~(\ref{eq:00<Pab>})), which is independent of both the neutrino
path length $L$ and the neutrino energy $E$. The effects of both
neutrino oscillations and neutrino decoherence are said to be
completely washed out.

\section{Equivalence between the Two Models}
\label{sec:equiv}

In the previous section, we have seen that the two formalisms,
neutrino oscillations and neutrino decoherence, have the same
dependence for the transition probabilities. In both scenarios, the
oscillation factor $\cos \frac{\Delta m^2 L}{2E}$ is damped by an
exponential factor ($e^{-2 \sigma^2 \left( \Delta m^2 \right)^2}$ or
$e^{-d^2 L}$).

Comparing Eqs.~(\ref{eq:neuosc}) and (\ref{eq:neudeco}) with each
other, we obtain
\begin{equation}
2 \sigma^2 \left(\Delta m^2\right)^2 = d^2 L,
\end{equation}
{\it i.e.}, the decoherence parameter $d$ is related to the damping
parameter $\sigma$ as
\begin{equation}
d = \frac{\sqrt{2} \Delta m^2}{\sqrt{L}} \sigma,
\label{eq:ds}
\end{equation}
which means that the two investigated models must be equivalent if
this condition is fulfilled. The units of the decoherence parameter
$d$ is $[d] = {\rm eV}^{1/2}$. This investigation could of course be
extended to $n$ neutrino flavors, but we will not do this here.

Next, we will now try to estimate the decoherence term ${\cal
D}[\rho]$ (or the decoherence parameter $d$).
The damping parameter $\sigma$ can be written as
\begin{equation}
\sigma = \frac{L}{4E} r, \quad \mbox{where $r = \frac{\Delta L}{L} +
\frac{\Delta E}{E}$ (pessimistic) or $r = \sqrt{\left(\frac{\Delta
L}{L}\right)^2 + \left(\frac{\Delta E}{E}\right)^2}$ (optimistic)}.
\label{eq:sr}
\end{equation}
Inserting Eq.~(\ref{eq:sr}) into Eq.~(\ref{eq:ds}) yields
\begin{equation}
d = \frac{\Delta m^2 \sqrt{L}}{2 \sqrt{2} E} r
\end{equation}
or
\begin{equation}
d^2 = \frac{\left(\Delta m^2\right)^2 L}{8 E^2} r^2.
\end{equation}
Thus, we obtain the following estimate for the decoherence term
\begin{equation}
{\cal D} \propto d^2 \sim \frac{\left(\Delta m^2\right)^2 L}{E^2}
\end{equation}
if we assume that $r^2 = {\cal O}(1)$.

Recently, other estimates of the decoherence term ${\cal D}[\rho]$ have been
found by Lisi {\it et al.} \cite{lisi00} and Adler
\cite{adle00}. Their different estimates are
\begin{equation}
{\cal D} \sim \frac{H_m^2}{M_{{\rm Planck}}} \sim \frac{E^2}{M_{{\rm Planck}}}
\end{equation}
and
\begin{equation}
{\cal D} \sim \frac{(\Delta H_m)^2}{M_{{\rm Planck}}} \sim
\frac{\left(\Delta m^2\right)^2}{E^2 M_{{\rm Planck}}},
\end{equation} 
respectively, where again ${\cal D} \propto d^2$, $H_m \simeq E$,
$\Delta H_m \simeq \frac{\Delta m^2}{2E} = k$, and $M_{{\rm Planck}}$
is the Planck mass scale. They argue that the decoherence could be
due to {\it e.g.} quantum gravity. We, on the other hand, have here argued
more phenomenologically.

In the next section, we are going to estimate the decoherence
parameter for atmospheric neutrinos.

\section{Estimation of the decoherence parameter for atmospheric neutrinos}
\label{sec:est}

The path length for atmospheric neutrinos, which is easily obtained
from geometrical considerations, is given by
\begin{equation}
L \equiv L(\cos \vartheta) = \sqrt{R_\oplus^2 \cos^2 \vartheta + 2 R_\oplus
d + d^2} - R_\oplus \cos \vartheta,
\end{equation}
where $\vartheta$ is the zenith angle, $R_\oplus$ is the radius of the
Earth ($R_\oplus \simeq 6400 \, {\rm km}$), and $d$ is the typical
altitude of the production point of atmospheric neutrinos above the
surface of the Earth ($d \simeq 10 \, {\rm km}$). The uncertainty in
the path length is mainly determined by $\Delta \cos \vartheta$ and therefore
\begin{equation}
\Delta L = \left\vert \frac{\partial L(\cos \vartheta)}{\partial \cos
\vartheta} \right\vert \Delta \cos \vartheta = \frac{R_\oplus L}{L +
R_\oplus \cos \vartheta} \Delta \cos \vartheta, \quad \mbox{{\it i.e.},} \quad
\frac{\Delta L}{L} = \frac{R_\oplus}{\sqrt{R_\oplus^2 \cos^2 \vartheta
+ 2 R_\oplus d + d^2}} \Delta \cos \vartheta.
\label{eq:uncL}
\end{equation}
The uncertainty in the neutrino energy for the Super-Kamiokande
experiment \cite{tosh00} as well as for the future MONOLITH experiment
\cite{geis00} is of the order of magnitude
\begin{equation}
\Delta E \sim E, \quad \mbox{{\it i.e.},} \quad \frac{\Delta E}{E} =
{\cal O}(1),
\label{eq:uncE}
\end{equation} 
even though the energy resolution is better for the MONOLITH than
Super-Kamiokande they are of the same order of magnitude.

Thus, using Eqs.~(\ref{eq:sigma}), (\ref{eq:uncL}), and
(\ref{eq:uncE}) as well as inserting numerical values, we
have the following upper bound for the damping parameter $\sigma_{\rm
atm.}$ for up-going ($\cos \vartheta = -0.95$ and $\Delta \cos
\vartheta = 0.1$ \quad $\Rightarrow$ \quad $L \simeq 12000 \, {\rm km}$ and
$\Delta L/L \simeq 0.11$) atmospheric neutrinos
\begin{equation}
\sigma_{\rm atm.} \lesssim 3.0 \cdot 10^{-4} \, {\rm m/eV} \simeq
6.0 \cdot 10^{-11} \, {\rm m^2} \simeq 1.5 \cdot 10^3 \, {\rm eV}^{-2}.
\end{equation}

From two flavor neutrino oscillations analyses of Super-Kamiokande
data, assuming $\nu_\mu$-$\nu_\tau$ oscillations, the atmospheric mass
squared difference $\Delta m_{\rm atm.}^2$ has been measured to be
$\Delta m_{\rm atm.}^2 \simeq 3.2 \cdot 10^{-3} \, {\rm eV}^2$
(and the mixing angle $\theta_{\rm atm.} \simeq 45^\circ$)
\cite{tosh00}. This means that the decoherence parameter for
atmospheric neutrinos is
\begin{equation}
d_{\rm atm.} = \frac{\sqrt{2} \Delta m_{\rm atm.}^2}{\sqrt{L(\cos
\vartheta = -0.95)}} \sigma_{\rm atm.} \lesssim 2.8 \cdot 10^{-8} \,
{\rm eV}^{1/2}
\end{equation}
or
\begin{equation}
d_{\rm atm.}^2 \lesssim 7.9 \cdot 10^{-16} \, {\rm eV} = 7.9 \cdot
10^{-25} \, {\rm GeV} \sim 10^{-24} \, {\rm GeV}.
\end{equation}

Recently, Lisi {\it et al.} have discussed three scenarios for the
decoherence parameter on the form $d_{\rm atm.}^2 = \gamma_0 \left(
\frac{E}{{\rm GeV}} \right)^n$, where $\gamma_0$ is a constant and $n
= -1,0,2$ \cite{lisi00}.
Our result is in general agreement with the ones obtained by Lisi
{\it et al.} ({\it i.e.}, they are all comparable within some orders
of magnitude), which are $d_{\rm atm.}^2 < 0.9 \cdot 10^{-27}
\, {\rm GeV}$ at 90\% C.L. ($n = 2$), $d_{\rm atm.}^2 < 3.5 \cdot
10^{-23} \, {\rm GeV}$ at 90\% C.L. ($n = 0$), $d_{\rm atm.}^2 < 4.1
\cdot 10^{-23} \, {\rm GeV}$ at 95\% C.L. ($n = 0$), $d_{\rm atm.}^2 <
5.5 \cdot 10^{-23} \, {\rm GeV}$ at 99\% C.L. ($n = 0$), and $d_{\rm
atm.}^2 < 2 \cdot 10^{-21} \, {\rm GeV}$ at 90\% ($n = -1$)
\cite{lisi00}. They found that $d_{\rm atm.}^2 \propto E^{-1}$ is
favored, which is natural from our point of view, since this would
mean that the dependence of the exponentially damping factor in the
transition probabilities is on the form $e^{-\alpha L/E}$, where
$\alpha$ is some constant, {\it i.e.}, it is exponentially decreasing
with the $L/E$ dependence. They also fitted the data with a pure
neutrino decoherence model ($\Delta m_{\rm atm.}^2 = 0$) and obtained
$d_{\rm atm.}^2 = 1.2 \cdot 10^{-21} \, {\rm GeV}$ ($n = -1$)
\cite{lisi00}. In all their fits, they obtained maximal mixing, {\it
i.e.}, $\theta_{\rm atm.} = 45^\circ$, as their best fit value.

\section{Summary and conclusions}
\label{sec:sc}

In conclusion, we have shown, in this paper, that a Gaussian averaged
neutrino oscillation model and a neutrino decoherence model are
equivalent if Eq.~(\ref{eq:ds}) is satisfied, {\it i.e.}, $d =
\frac{\sqrt{2} \Delta m^2}{\sqrt{L}} \sigma$. Our upper bound on
$d_{\rm atm.}^2 \lesssim 10^{-24} \, {\rm GeV}$ then immediately
follows from the results of the Super-Kamiokande analyses, {\it i.e.},
the neutrino decoherence model is just a reparameterization of the
Gaussian averaged neutrino oscillation model or vice versa. We have
also estimated the decoherence term to be ${\cal D} \propto d^2 \sim
\frac{\left(\Delta m^2\right)^2 L}{E^2}$, which is different from
earlier results including the Planck mass scale $M_{{\rm Planck}}$
and motivations that decoherence could be induced by new physics
beyond the Standard Model such as {\it e.g.} quantum gravity.

\begin{acknowledgments}
I would like to thank Robert Buras, Martin Freund, G{\"o}ran Lindblad, Manfred
Lindner, Serguey T. Petcov, Georg G. Raffelt, Dmitri V. Semikoz, H{\aa}kan
Snellman, and Leo Stodolsky for useful discussions.

This work was supported by the Swedish Foundation for International
Cooperation in Research and Higher Education (STINT) %,
%the Wenner-Gren Foundations,
and the ``Sonderforschungsbereich 375 f{\"u}r
Astro-Teilchenphysik der Deutschen Forschungsgemeinschaft''.
\end{acknowledgments}

% Create the reference section using BibTeX:
\bibliography{references_r}

\end{document}